# Stakeholder's interdependencies and their role in sustainable business model innovation

## Iqra S. Khan*

University of Oulu, Pentti Kaiteran katu 1, 90014 Oulu, Finland.
E-mail: iqra.khan@oulu.fi
* Corresponding author

## Jukka Majava

University of Oulu, Pentti Kaiteran katu 1, 90014 Oulu, Finland.
E-mail: jukka.majava@oulu.fi

**Abstract:** Sustainable innovation requires in-time development, diversification and transformation of business models from one to another. Business model innovation, development and transformation for sustainability incorporates economic, environmental and social value by advancing the management of the stakeholders into the business model. Except a little research on business model inter-dependencies, scant research has been done on stakeholder's inter-dependencies in order to understand their nature and relationship while developing or transforming a business model and creating an impact on environment, society and economy. Therefore, current research uses actor dependency model to analyze four different kind of inter-dependencies, namely, goal-dependency, task-dependency, resource-dependency and soft-goal dependency. The ecology of business model experimentation map is used as a tool for practical understanding of sustainable business modelling with a multi-actor approach in a workshop setting. The findings will help to understand how stakeholders depend on each other while developing a business model for sustainability and innovation.

**Keywords:** Sustainable business models; Business model for sustainability; innovation; stakeholder dependencies; workshop; joint value

## 1   Introduction

Sustainable business models in scientific research, businesses and policy-making are considered significant to yield a positive environmental and societal impact (Ludeke-Freund and Dembek, 2017; Bocken, Boons, and Baldassarre, 2019). Yet, academic scholars had accumulated little insight into how stakeholder's interdependencies as a firm-level antecedent and a significant business model component influences business model innovation for sustainability (Adams et al., 2015; Bstieler et al., 2018). Stakeholders in sustainable business model innovation involves the principal organization and other stakeholders such as customers, suppliers and partners/employees and government who shape and transform the new business models for sustainability in cooperation (Bocken et al., 2013) thus creating a shared value for the system.

Innovation in this form focus on collaborative learning and designing rather than a managerial-centric design form a focal organization (Roome and Lauche, 2016). However, generating shared value between stakeholders is difficult to comprehend in practice. It first needs the specification of system boundaries as to know who is centrally involved in the development of business model for sustainability (Bocken et al., 2019). Such centralization of the actors is deeply related with the notion of density. Density measures the extent to which everyone is connected to each other and so is defined as the ratio of existing ties to the theoretical maximum (Knoke, Wasserman and Faust, 1996). Thus a highly stakeholder dense environment means high stakeholder's connectivity to collaborate, share information and forming coalitions resulting into significant developments.

Sustainability-focused value-creation and innovation is known as criteria firms use to develop joint social and business value by integrating multiple societal and environmental factors to new product development (Bstieler et al., 2018). Such integrations create managerial and strategic contradictions in terms of different context-dependencies between actors, business models and technologies. These contradictions raise problem uncertainties and thus signify a growing need to consider the development of sustainable business models with high stakeholder density (Hämäläinen, 2015). Therefore, extant research have called to investigate sources and antecedents of sustainable innovation (Adams, Jeanrenaud, Bessant, Denyer and Overy, 2015), specifically stakeholder and firm related sources and challenges to sustainable business model innovation for new product development (Bstieler et al., 2018). Prior research have typically focused on the benefits of sustainability or on different institutional and macro level factors affecting firms' sustainable practices (Ludeke-Freund and Dembek, 2017; Hämäläinen, 2015) and more recently on business model dependencies (Bocken et al., 2019). However, a comprehensive understanding of the stakeholder interdependencies as a significant unit of business modelling to sustainable innovation pleads further attention (Adams et al., 2015; Bstieler et al., 2018). Therefore, current study uses a theoretical lens of business model thinking to explore stakeholder inter-dependencies, their probable nature and investigates if the stakeholders create any further opportunity, value or competitive advantage for each other in sustainable business model innovation.

*Research Questions*

RQ1: What is the nature of interdependencies between stakeholders in the development of sustainable business models?
RQ2: How do stakeholder's interdependencies influence sustainable business model innovation in a highly stakeholder dense environment?
RQ3: How do stakeholders create opportunity, value and advantage for each other while developing sustainable business models?

## 2   Current understanding:

In the present hit and hype of industry 4.0 it is important to understand the challenges of business modelling, value creation, allocation of resources in the context of open dynamic system of networks, capabilities and ever-changing social, economic and environmental factors. To such challenges, sustainable business models positioned towards circular economies are considered as the key to value propositions (Ludeke-Freund and Dembek, 2017). Sustainable business models are considered as the product of important concepts such as business models, stakeholder management, strategic and sustainable value creation (Geissdoerfer et al., 2016). They aim to improve overall effectiveness of the business from economic, environment and social perspectives. These developments can be attained by managing the stakeholders with a long-term strategy (Geissdoerfer et al., 2016).

However, sustainable value creation and innovation demands major redesigns in business models (Boons and Ludeke-Freund, 2013) and necessitates various new viewpoints on value and stakeholders (Ludeke-Freund and Dembek, 2017). To redesign, sustainable business model literature signifies the role of the network partners (Boons and Ludeke-Freund, 2013). This reflects that the context of sustainable business model is composed of dynamic system of stakeholders and manifold knowledge creation processes. However, with the involvement of new actors as partners, the process of business modelling turns into a new multi-layered structure of inter-organizational engagements and network activities resulting into external dependencies. This negates the process of business modelling as a hierarchical activity and implies that business modelling is a co-evolutionary process that changes and evolves according to new configurations of dependencies developed through networking (Oskam, Bossink and De Man, 2018). These dynamic stakeholders interactions necessitates the need to explore the nature of their interdependencies and how they create opportunity, value and advantage for each other to develop further sustainable businesses and innovations.

One way or another, all the organizations depend on others/stakeholders for some goals to be accomplished, duties/tasks to be executed and resources to be well appointed. This reflects that the more closer/denser stakeholders work with and for each other the more crucial the examination of the stakeholder's dependencies become. Considering goals, activities and resources as shared responsibilities of an organization and related stakeholders, this study aims to use actor dependency model (Yu and Mylopoulos, 1995) in order to access the nature of stakeholder's inter-dependencies in business modelling for sustainable innovation. This model entails four types of principle dependencies: 1) Goal dependency (dependency to achieve the goal), Task dependency (dependency to complete an activity), resource dependency (dependency based on some physical or informational entity) and soft-goal dependency (a twin of goal and task dependency where stakeholders perform some tasks to reach the soft-goal). Actor dependency model will help to understand the nature of stakeholder's inter-dependencies in the development of sustainability-focused business models.

## 3 Research design:

Research methods for this study will aim to build upon the existing expertise and practices to drive the growth of local businesses specifically small and medium size manufacturing enterprises (SMEs) located in northern part of Finland. For this purpose, action research will be used as it fulfills the basic principle of action research, which is "relevance" in terms of having stakeholders as the "actors" to be assisted for improving their actions. Furthermore, "there is a dual commitment in action research to study a system and concurrently to collaborate with members of the system in changing it, which is together regarded as a desirable direction. Accomplishing this twin goal requires the active collaboration of researcher and participant, and thus it stresses the importance of co-learning as a primary aspect of the research process" (ABL group, 1997). This perfectly implies to the context of current research as it involves a system approach to address dependencies with the cooperation of certain networks as (university researchers, industrial experts) and stakeholders to develop sustainable value-creation and innovation.

More specifically, as proposed by (Bocken et al., 2019) the ecology of business model experimentation map will be used as a tool for generic and more practical understanding of business model ecologies with a multi-actor approach and a system-level perspective in a workshop setting. This workshop will be used as an analogy to participative action research. This is a form of visualized brainstorming. Aim of this exercise is to engage the participant's futures thinking as well as to understand the operational environment regarding how closely and in which ways they interact for sustainability and innovation. Putting the workshop method to our study context, workshop data will help us understand how a desired, future-oriented change in the firm's business model affect the sustainable business model within its stakeholder's network. Consequently, this will contribute to understand the interdependencies within the business model and investigate its implications in sustainable business model transformation/innovation process.

As a second iteration, conversational interviews will be conducted with the prospective stakeholders of the targeted companies. This will allow the researcher to embed and compare the individual views to the research findings. In addition, the research data will be supplemented periodically with online market research and possible ethnographic observations.

## 4 Findings:

In order to improve system-effectiveness and promote business models focused towards sustainability, the findings will help to do systematic examination of the stakeholder's networks and their inter-dependencies. In order to see organizational re-design towards sustainability, this study aims to use actor-dependency model (Yu and Mylopoulos, 1995) to map the stakeholder's relationships. This model takes goal-based dependencies, activity-based dependencies and resource-based dependencies needed as means to end in the organizations and related networks. Using an actor dependency model will first help to understand appropriate representation of an organizational configurations (business model elements, stakeholder positioning, and transformation towards sustainability). In addition, the findings will help to sought out stakeholder's major concerns about current and proposed business model configurations and subject matters such as power, values and conflicts in their relationships. In sum, the results will help to understand stakeholder's roles and their relationships in developing and transforming business models towards sustainability and innovation.

*Contribution:*

>The results will help in finding out the extent to which the stakeholders are connected and nature of their inter dependencies in SME's. In addition, the results will be a stepping-stone for the existing organizations to find new, joint sustainable practices or to adopt utterly new sustainable business models towards the transition of a comprehensive sustainable economic and innovative social organization. Results will further contribute to the opportunity advancement processes with innovation research that has encompassed research on business model innovation.

*Practical implications:*

>Results will help inter-linked organizations, stakeholders and managers to find operational and management patterns that can potentially enhance growth of their businesses and cause improvement in productivity. In addition, the findings will help develop new policies and recommendations for new manufacturing start-ups.

## 5  Areas for Feedback and Development:

Despite of my readings and knowledge on the topic, there is a need to get feedback (from someone who has more work and knowledge on the topic already) on what literature streams, concepts and constructs are needed to build a narrative and background understanding that leads to this research question and problem. For instance, what could be the best explanatory theme as an underlying base for current research: Business model research or Stakeholder theory/management research?

Regarding the research design, it is crucial that it matches the research question and phenomenon, so it is important to consider, what kind of empirical research design would be the best for this research phenomenon considering available resources and situation. Since the work is quite on early stage and the case companies are yet to be decided, guidance is needed about the design that will best inform this type of research. Action research methodology is seriously considered at the moment. Additionally, the case context is crucial and may influence the theoretical background, so that link is significant as well. For instance, the SME context might have different business model innovation processes and stakeholder dynamism compared to more established and larger corporations. It will be useful if a little bit more information is received about the domain and method theory for this research. This will help to reflect more on the methodology part afterwards. As for the initial feedback on the first draft submission, it was asked to provide construct clarity on few narratives used in the introduction. The comments are incorporated now, but it would be valuable to know as if the paper is reflecting any connectivity now?

Another thing would be a reflection on the research questions: if the research questions match with the details described in the introduction and do, they lead the authors to a valid problem statement. In addition, it would be valuable to know more about business model perspectives in discussion.